\documentclass[usenatbib,useAMS]{mn2e}
\usepackage{graphicx}
\usepackage{subfigure}

\usepackage{aas_macros}
\newcommand{\gta}{{\small\raisebox{-0.6ex}
{$\,\stackrel{\raisebox{-.2ex}{$\textstyle >$}}{\sim}\,$}}}
\newcommand{\lta}{{\small\raisebox{-0.6ex}
{$\,\stackrel{\raisebox{-.2ex}{$\textstyle <$}}{\sim}\,$}}}

\title[]{On the evolutionary status of short period cataclysmic variables}

\author[]{S.\,P.\ Littlefair$^{1}$, V.\,S.\, Dhillon$^{1}$, 
T.\,R.\, Marsh$^{2}$, B.\,T.\, G\"{a}nsicke$^{2}$, \newauthor John Southworth$^{2}$,
 I.\, Baraffe${^{3,4}}$,  C.\,A.\, Watson$^{1}$, C.\, Copperwheat${^2}$ \\
$^1$Dept of Physics and Astronomy, University of Sheffield, S3 7RH, UK \\
$^2$Dept of Physics, University of Warwick, Coventry, CV4 7AL, UK\\
$^3$Ecole Normale Sup\'{e}rieure de Lyon, CRAL, 46 all\'{e}e d'Italie,
69364 Lyon Cedex 07, CNRS UMR 5574, Universit\'{e} de Lyon 1, France\\
$^4$Max-Planck-Institut fuer Astrophysik, Postfach, 1317, 85741 Garching, Germany\\}

\date{\center{\Large Submitted for publication in the Monthly Notices of the
Royal Astronomical Society \\
\vspace{.5cm} \today}}

\begin{document}
\maketitle

\begin{abstract} 
We present high-speed, three-colour photometry of seven short period (P$_{\rm orb} \le 95$ mins) eclipsing CVs from the Sloan Digital Sky Survey. We determine the system parameters via a parametrized model of the eclipse fitted to the observed lightcurve by $\chi^2$ minimization.  Three out of seven of the systems possess brown dwarf donor stars and are believed to have evolved past the orbital period minimum. This is in line with predictions that 40--70 per cent of CVs should have evolved past the orbital period minimum. Therefore, the main result of our study is that the missing population of post-period minimum CVs has finally been identified. The donor star masses and radii are, however, inconsistent with model predictions; the donor stars are approximately 10 per cent larger than expected across the mass range studied here. One explanation for the discrepancy is enhanced angular momentum loss (e.g. from circumbinary discs), however the mass-transfer rates, as deduced from white dwarf effective temperatures, are not consistent with enhanced angular momentum loss. We show it is possible to explain the large donor radii without invoking enhanced angular momentum loss by a combination of geometrical deformation and the effects of starspots due to strong rotation and expected magnetic activity.  Choosing unambiguously between these different solutions will require independent estimates of the mass-transfer rates in short period CVs. 

The white dwarfs in our sample show a strong tendency towards high masses. We show that this is unlikely to be due to selection effects. The dominance of high-mass white dwarfs in our sample implies that erosion of the white dwarf during nova outbursts must be negligible, or even that white dwarfs grow in mass through the nova cycle. Amongst our sample there are no Helium core white dwarfs, despite predictions that 30--80 per cent of short period CVs should contain Helium core white dwarfs. We are unable to rule out selection effects as the cause of this discrepancy. 
\end{abstract} 


\section{Introduction}
\label{sec:introduction}
Cataclysmic variable stars (CVs) are a class of interacting binary system undergoing mass-transfer, usually via a gas stream and accretion disc, from a Roche-lobe filling secondary to a white dwarf primary. A bright spot is formed at the intersection of the disc and gas stream, giving rise to an `orbital hump' in the lightcurve at phases $0.6-1.0$ due to foreshortening of the bright-spot. For a basic review of CVs, see \cite{warner95a}. The secular evolution of CVs represents a long-standing problem in astrophysics, with wide ranging implications for all close binary systems \cite[see][for example]{kolb93}. Close binary star evolution is driven by angular momentum loss, sustaining mass-transfer which in turn results in changes in the binary's orbital period. Since the orbital period is easy to determine,  it is accurately known for many CVs, and the aim of CV evolutionary theory has been to reproduce the observed period distribution. Salient features which must be explained  are the period gap -- a dearth of systems with periods between 2 and 3 hours, and the  period minimum -- a sharp cut-off in the period distribution at 76.2 mins \citep{knigge06}. The period gap is thought to be caused by a sudden reduction in angular momentum loss when the donor star becomes fully convective. The cause of this reduction is the cessation of magnetic braking \citep[e.g][]{robinson81}. Whilst observational evidence for a dramatic change in magnetic braking rate at the convective boundary exists \citep{delfosse98,reiners07}, both the theoretical justiÞcation \citep{tout92} and the observational basis \citep{andronov03} for the so-called disrupted magnetic braking model have been challenged. Despite this, disrupted magnetic braking remains the best explanation to date for the period gapÕs existence. 

The period minimum is caused by the response of the donor star to ongoing mass loss \citep[e.g][]{paczynski81}. Typically, the donor shrinks in response to mass loss, causing the orbital period to decrease. When the donor nears 
the substellar limit, however, two effects come into play. Firstly, the thermal timescale  of the donor exceeds the mass-transfer timescale, and the donor cannot shrink rapidly  enough in response to mass loss. Secondly, for substellar donors, changes in the internal structure mean that the donor expands in response to mass loss. Both these effects mean that continuing mass loss results in an increasing orbital period. Thus there exists a period minimum, the exact value of which depends upon the interplay between the thermal timescale of the donor, and the mass-transfer rate. Furthermore, since the number density of CVs, at a given period, scales inversely with the rate at which its period evolves, a significant ``spike'' of systems is expected to accumulate at the minimum period.

Significant problems exist with the explanation of the period minimum. Firstly, the observed minimum period is around 10 minutes longer than the predicted value \cite[see][for a review]{kolb02}. Also, the predicted period spike is not seen in the orbital period distribution. Moreover, whilst significant numbers of CVs should have passed the period minimum, with the exact figure ranging from around 70 per cent \citep{kolb93,howell97} to nearer 40 per cent \citep{willems05}. Despite this, it has proved very hard to identify these so-called Òpost-bounceÓ systems \citep{littlefair03,patterson05}. Until recently, it was not known whether this reflected a genuine absence of post-bounce systems, or a selection effect, caused by the insensitivity of CV surveys to low mass-transfer rate systems \citep{kolb99}, and the   difficulty of observing very low mass  donor stars against the backdrop of the accretion disc and white dwarf \citep{littlefair03}.  Recent developments have allowed both these shortcomings to be overcome. The  Sloan Digital Sky Survey (SDSS) \citep{york00} goes much fainter than previous surveys, and as objects are selected on the basis of their spectra, CVs need not show outbursts  to be included. The SDSS sample could therefore contain a large number of post-period minimum systems. Furthermore, the availability of {\sc ultracam} \citep{dhillon07} means that it is possible to obtain high quality lightcurves of the eclipses of faint CVs within the SDSS. This is significant because the lightcurves of eclipsing CVs allow the system parameters to be determined to a high degree of precision \citep{wood86a}, even when the donor star is not directly visible. Taking advantage of this technique, \cite{littlefair06} recently presented the first robust identiÞcation of a post-bounce CV. Here we expand upon those results, by determining the system parameters of a small sample of SDSS CVs with orbital periods below 95 minutes.

Our sample consists of seven systems discovered in the SDSS.  The systems selected are SDSS J090350.73+330036.1, SDSS J103533.03+055158.4, SDSS J122740.83+513925.0, SDSS J143317.78+101123.3, SDSS J150137.22+550123.3, SDSS J150240.98+333423.9 and SDSS J150722.30+523039.8 (hereafter SDSS 0903, 1035, 127, 1433, 1501, 1502 and 1507, respectively). The systems were flagged as high inclination by the presence of broad, double peaked, emission lines and were subsequently found to be eclipsing by follow-up observations. SDSS 1433, 1501, 1502 and 1507 were reported as eclipsing in the discovery papers \citep{szkody04,szkody05,szkody06,szkody07}. SDSS 1035 was found to be eclipsing from time-resolved spectroscopy \citep{southworth06}. Amateur observations in outburst revealed SDSS 1227 to be eclipsing \citep{shears07}, whilst \cite{dillon08} discovered eclipses in SDSS 0903. Our sample consists of all the eclipsing dwarf novae within the SDSS, with orbital periods below 95 minutes, which were known of and visible at the time of observations. As such, selection effects within our sample should be minimal. Analysis of, and results from, SDSS 1035 and SDSS 1507 are presented in \cite{littlefair06} and \cite{littlefair07}, respectively. In this paper, we present the analysis of the remaining systems, and discuss the implications of our results for the evolution of short period cataclysmic variables.

\section{Observations}
\label{sec:obs}
\begin{table*}
\begin{center}
\caption[]{Journal of observations. The deadtime between exposures is 0.024 seconds, and the
  relative GPS time-stamping on each data point is accurate to 50
  $\mu$s.}
\begin{tabular}{@{\extracolsep{-1.25mm}}ccrrcccccc}
\hline
Date & Object & Start Phase & End Phase  & BMJD$_{\rm eclipse}$ &t$_{\rm exp}$ (s) & Num Points & Seeing (arcsec) & Airmass & Photometric?\\ \hline
2006 Mar 05 & SDSS 0903 & -0.90 & 0.20     & 53799.894700(6) & 3.990 & 1396 & 1.3--3.8 & 1.084--1.349 & Y\\
2006 Mar 05 & SDSS 0903 & 1.85 & 2.16      & 53800.012854(6) & 3.990 & 395 & 1.3--6.5 & 1.012--1.031 & Y\\
2006 Mar 07 & SDSS 0903 & 33.90 & 34.32 & 53801.903198(6) & 3.990 & 532 & 0.7--1.3 & 1.042--1.092 & Y\\
2006 Mar 07 & SDSS 0903 & 34.95 & 35.27 & 53801.962279(6) & 3.990 & 410 & 0.8--1.4 & 1.003--1.005 & Y\\
2006 Mar 07 & SDSS 0903 & 35.85 & 36.22 & 53802.021361(6) & 3.990 & 477 & 0.9--1.5 & 1.026--1.061 & Y\\
2006 Mar 07 & SDSS 0903 & 36.81 & 37.19 & 53802.080442(6) & 3.990 & 481 & 0.9--1.4 & 1.155--1.247 & Y\\
2006 Mar 07 & SDSS 0903 & 37.80 & 38.19 & 53802.139501(6) & 3.990 & 493 & 1.1--2.5 & 1.482--1.712 & Y\\
2006 Mar 08 & SDSS 0903 & 49.77 & 50.13 & 53802.848386(6) & 3.990 & 463 & 0.8--1.6 & 1.225--1.339 & Y\\
2006 Mar 08 & SDSS 0903 & 50.78 & 51.18 & 53802.907462(6) & 3.990 & 509 & 0.9--1.8 & 1.043--1.092 & Y\\
2006 Mar 08 & SDSS 0903 & 51.90 & 52.16 & 53802.966529(6) & 3.990 & 330 & 0.9--1.5 & 1.003--1.004 & Y\\
2006 Mar 08 & SDSS 0903 & 52.73 & 53.20 & 53803.025601(6) & 3.990 & 599 & 1.1--3.3 & 1.026--1.073 & Y\\
\hline
2006 Mar 04 & SDSS 1035 & -0.17 & 0.17    & 53798.98148(1) & 3.980 & 413 & 1.3--3.6 & 1.150--1.207 & Y\\
2006 Mar 04 & SDSS 1035 & 0.68 & 1.25     & 53799.03848(2) & 3.980 & 708 & 1.3--2.5 & 1.086--1.099 & Y\\
2006 Mar 04 & SDSS 1035 & 1.85 & 2.13     & 53799.09548(2) & 3.980 & 344 & 1.3--2.5 & 1.121--1.155 & Y\\
2006 Mar 05 & SDSS 1035 & 18.87 & 19.16 & 53800.06459(2) & 3.980 & 354 & 1.5--6.8 & 1.090--1.105 & Y\\
2006 Mar 07 & SDSS 1035 & 50.83 & 51.12 & 53801.88881(2) & 3.980 & 363 & 1.0--1.7 & 1.675--1.913 & Y\\
2006 Mar 07 & SDSS 1035 & 51.64 & 52.14 & 53801.94582(2) & 3.980 & 620 & 1.2--2.1 & 1.247--1.406 & Y\\
2006 Mar 07 & SDSS 1035 & 52.67 & 53.14 & 53802.00282(2) & 3.980 & 582 & 0.7--1.5 & 1.100--1.146 & Y\\
2006 Mar 08 & SDSS 1035 & 71.72 & 72.13 & 53803.08596(2) & 3.980 & 498 & 0.9--2.1 & 1.111--1.158 & Y\\
\hline
2006 Mar 01 & SDSS 1227 & -0.31 &  0.71        & 53796.2482445(7) & 3.500 & 1579 & 1.1--1.9 & 1.220--1.515 & Y\\
2006 Mar 02 & SDSS 1227 & 15.27 & 16.42      & 53797.2554528(8) & 2.990 & 2084 & 0.8--2.0 & 1.177--1.463 & Y\\
2006 Mar 10 & SDSS 1227 & 139.92 & 140.11 & 53805.0613010(9) & 3.500 & 290 & 1.0--1.8 & 1.103--1.116 & Y\\
\hline
2007 Jun 10 & SDSS 1433 & 7443.91 & 7445.24 & 54262.12450(1) & 1.990 & 3133 & 0.5--1.2 & 1.245--1.574 & N\\
2007 Jun 16 & SDSS 1433 & 7551.70 & 7552.21 & 54262.17874(1) & 1.990 & 1201 & 0.7--2.3 & 1.387--1.569 & N\\
2007 Jun 16 & SDSS 1433 & 7554.71 & 7555.25 & 54264.13139(1) & 1.990 & 1282 & 0.8--2.1 & 1.315--1.457 & N\\
2007 Jun 21 & SDSS 1433 & 7646.62 & 7647.78 & 54273.13537(1) & 1.990 & 2710 & 0.8--1.8 & 1.313--1.742 & Y\\
2007 Jun 21 & SDSS 1433 & 7647.81 & 7648.18 & 54273.18963(1) & 1.990 & 876 & 1.2--2.0 & 1.761--2.068 & Y\\
\hline
2006 Mar 04 & SDSS 1501 & -0.66 & 0.16    & 53799.211567(6) & 4.985 & 812 & 1.2--3.3 & 1.115--1.149 & Y\\
2006 Mar 05 & SDSS 1501 & 15.15 & 16.19 & 53800.121048(7) & 5.985 & 852 & 1.6--6.4 & 1.226--1.455 & Y\\
2006 Mar 07 & SDSS 1501 & 49.71 & 50.19 & 53802.053634(6) & 4.985 & 479 & 0.7--1.7 & 1.472--1.656 & Y\\
2006 Mar 07 & SDSS 1501 & 50.68 & 51.21 & 53802.110471(7) & 4.985 & 524 & 0.8--1.3 & 1.236--1.338 & Y\\
2006 Mar 07 & SDSS 1501 & 52.81 & 53.21 & 53802.224169(7) & 4.985 & 396 & 0.8--1.2 & 1.114--1.118 & Y\\
2006 Mar 08 & SDSS 1501 & 68.83 & 69.14 & 53803.133609(6) & 4.960 & 314 & 0.9--1.4 & 1.185--1.226 & Y\\
2006 Mar 08 & SDSS 1501 & 69.94 & 70.14 & 53803.190452(7) & 4.960 & 196 & 0.9--1.4 & 1.118--1.125 & Y\\
2006 Mar 08 & SDSS 1501 & 70.91 & 71.14 & 53803.247311(6) & 4.960 & 226 & 0.2--1.0 & 1.123--1.134 & Y\\
\hline
2006 Mar 04 & SDSS 1502 & -0.60 & 0.21     & 53799.140618(4) & 1.994 & 2068 & 1.1--3.7 & 1.091--1.266 & Y\\
2006 Mar 04 & SDSS 1502 & 1.80 & 2.12      & 53799.258414(7) & 3.984 & 418 & 1.4--2.0 & 1.009--1.025 & Y\\
2006 Mar 05 & SDSS 1502 & 16.88 & 17.16 & 53800.142070(6) & 3.984 & 363 & 1.6--5.9 & 1.088--1.135 & Y\\
2006 Mar 05 & SDSS 1502 & 17.86 & 18.66 & 53800.200966(6) & 3.991 & 1028 & 1.2--8.4 & 1.003--1.019 & Y\\
2006 Mar 05 & SDSS 1502 & 18.75 & 19.44 & 53800.259901(6) & 3.990 & 881 & 1.2--6.2 & 1.009--1.059 & Y\\
2006 Mar 07 & SDSS 1502 & 51.72 & 52.15 & 53802.203911(2) & 1.745 & 1252 & 0.8--1.2 & 1.004--1.019 & Y\\
2006 Mar 08 & SDSS 1502 & 67.96 & 68.15 & 53803.146461(3) & 3.240 & 303 & 0.9--1.6 & 1.062--1.086 & Y\\
2006 Mar 08 & SDSS 1502 & 68.93 & 69.14 & 53803.205371(6) & 2.589 & 411 & 0.8--1.4 & 1.004--1.007 & Y\\
2006 Mar 08 & SDSS 1502 & 69.88 & 70.23 & 53803.264277(3) & 1.992 & 885 & 0.9--1.3 & 1.027--1.059 & Y\\

\hline
\end{tabular}
\label{journal}
\end{center}
\end{table*}

On nights between Mar 01$^{st}$ 2006 and Mar 10$^{th}$ 2006, SDSS 0903, 1227, 1501 and SDSS 1502 were observed simultaneously in the SDSS-$u'g'r'$ colour bands using {\sc ultracam} \citep{dhillon07} on the 4.2-m William Herschel Telescope (WHT) on La Palma. On nights between 2007 Jun 10$^{th}$ and 2007 Jun 21$^{st}$, SDSS 1433 was observed simultaneously in the SDSS-$u'g'r'$ colour bands using {\sc ultracam} on the 8.2-m Very Large Telescope (VLT) in Chile. A complete journal of observations is shown in table~\ref{journal}.  The observations of SDSS 1035 are also included in table~\ref{journal}, as this information was not provided by \cite{littlefair06} due to space constraints. Data reduction was carried out in a standard manner using the {\sc ultracam} pipeline reduction software, as described in \cite{feline04a}, and a nearby comparison star was used to correct the data for transparency variations. Observations of a standard star taken at the start and the end of the night were used to correct the magnitudes to a standard system \citep{smith02}.  Because of the absence of a comparison star which was sufficiently bright in the $u'$-band, the $u'$-band data for SDSS 1507 was corrected using the $g'$-band data for the comparison star, with appropriate corrections for the magnitude difference and atmospheric extinction.

\section{Results}
\label{sec:results}

\subsection{Orbital Ephemerides}
\label{subsec:ephem}
\begin{table}
\begin{center}
\caption[]{Orbital Ephemerides}
\begin{tabular}{crr}
\hline
Object &  T$_0$ (BMJD) & \multicolumn{1}{c}{P$_{\rm orb}$ (days)} \\
\hline
SDSS 0903   &  53799.894707(2)    &  0.059073543(9) \\
SDSS 1035   &  53798.981469(8)    &  0.0570067(2) \\
SDSS 1227   &  53796.2482451(5)  &  0.062959041(7) \\
SDSS 1433   &  53858.35689(2)  &  0.054240679(2) \\
SDSS 1501   & 53799.211577(7)  &  0.0568412(2) \\
SDSS 1502   & 53799.140607(3)  &  0.05890961(5) \\
SDSS 1507   & 53798.239587(3)      &  0.04625828(4) \\
\hline
\end{tabular}
\label{eclipse_times}
\end{center}
\end{table}

The times of white dwarf mid-ingress $T_{\rmn{wi}}$ and mid-egress $T_{\rmn{we}}$ were determined by locating the minimum and maximum times, respectively, of the lightcurve derivative.  Mid-eclipse times, $T_{\rmn{mid}}$, were determined by assuming the white dwarf eclipse to be symmetric around phase zero and taking $T_{\rmn{mid}}=(T_{\rmn{we}}+T_{\rmn{wi}})/2$.  
Mid-eclipse times are presented in table~\ref{journal}. For SDSS 0903, SDSS 1227 and SDSS1433 we also included the eclipse times of \cite{dillon08}, \cite{shears07} and \cite{szkody07}, respectively. In each case, the ephemeris determined from our data was sufficient to project to the additional eclipse times without cycle ambiguity. The errors on all mid-eclipse times were adjusted to give $\chi^{2} = 1$, with respect to a linear fit.  Where two sources of eclipse times were used, we first set the errors on our own measurements by ensuring $\chi^{2} = 1$ with respect to a linear fit to our points alone, and then adjusted the error bars on the second set of eclipse times to give $\chi^{2} = 1$ with respect to a linear fit to all data points. From our data, only the $g'$-band and $r'$-band lightcurves were used, given the significantly poorer quality of the $u'$-band lightcurves. The results from the two colours were combined with a weighted mean to give the ephemerides shown in table~\ref{eclipse_times}.  There was no significant deviation from linearity in the $O-C$ times. The ephemerides in table~\ref{eclipse_times} were used to phase our data for the analysis which follows.

\subsection{System Parameters}
\label{sec:model}

The method used to determine system parameters has been described in detail by \cite{littlefair07}. Here we limit ourselves to a brief discussion of the method and its limitations.  We first assume the secondary star fills its Roche Lobe. The width of the white dwarf eclipse, $\Delta\phi$, then depends solely upon the inclination, $i$, and the mass ratio, $q$ \citep{bailey79}.  If we also assume that the gas stream follows a ballistic trajectory from the secondary, and that the bright spot lies along that ballistic trajectory, then the position of the bright spot depends upon $q$ and the outer radius of the accretion disc, $R_{\rm d}$. Combining the bright spot ingress and egress with the width of the white dwarf eclipse thus gives a system of three constraints ($\Delta\phi$ and the ingress and egress phases of the bright spot eclipse) and three unknowns ($q$, $i$ and $R_{\rm d}$) which can be solved to yield estimates of $q$, $i$, and $R_{\rm d}$. To get from the mass ratio to individual component masses, the white dwarf mass is determined from the white dwarf radius, assuming that the white dwarf follows a theoretical mass-radius relationship of appropriate effective temperature.  The white dwarf radius is itself measured from the duration of white dwarf ingress/egress, and the effective temperature estimated from the colours of white dwarf ingress/egress. Thus, determining the width of the white dwarf eclipse, the duration of white dwarf ingress/egress and the contact phases of the bright spot eclipse is sufficient to determine the component masses of the binary
system with minimal assumptions.

\cite{wood86a} showed how the above quantities can be measured from the derivative of the eclipse lightcurve, but in practice a physical model of the binary system is often fitted to the lightcurve instead. Although both methods give similar results, \cite{feline04} showed that model fitting gives a more robust determination of the system parameters in the presence of flickering than the derivative method. Both methods rely on the same three assumptions: that the bright spot lies on the ballistic trajectory from the secondary star, that the secondary star fills its Roche Lobe and that the white dwarf is accurately described by a theoretical mass-radius relation.  The systematic uncertainty introduced by the latter assumption can be estimated by comparing the results from different theoretical models and is shown by \cite{littlefair07} to be small compared to the statistical errors in lightcurve fitting. Whilst the assumption that the secondary fills its Roche Lobe is difficult to test directly, it is highly likely, given the presence of ongoing mass-transfer in these CVs. The assumption about the bright spot trajectory is also difficult to test, but relies upon the stellar orbits being circular, and the secondary star rotating synchronously. Both of these conditions are likely to be satisfied. Indeed, \cite{feline05} show that masses derived with this method are consistent with dynamical mass determinations in cataclysmic variables over a wide range of orbital periods, giving some confidence that our assumptions are correct.

\begin{figure*}
\begin{center}
\includegraphics[scale=0.75]{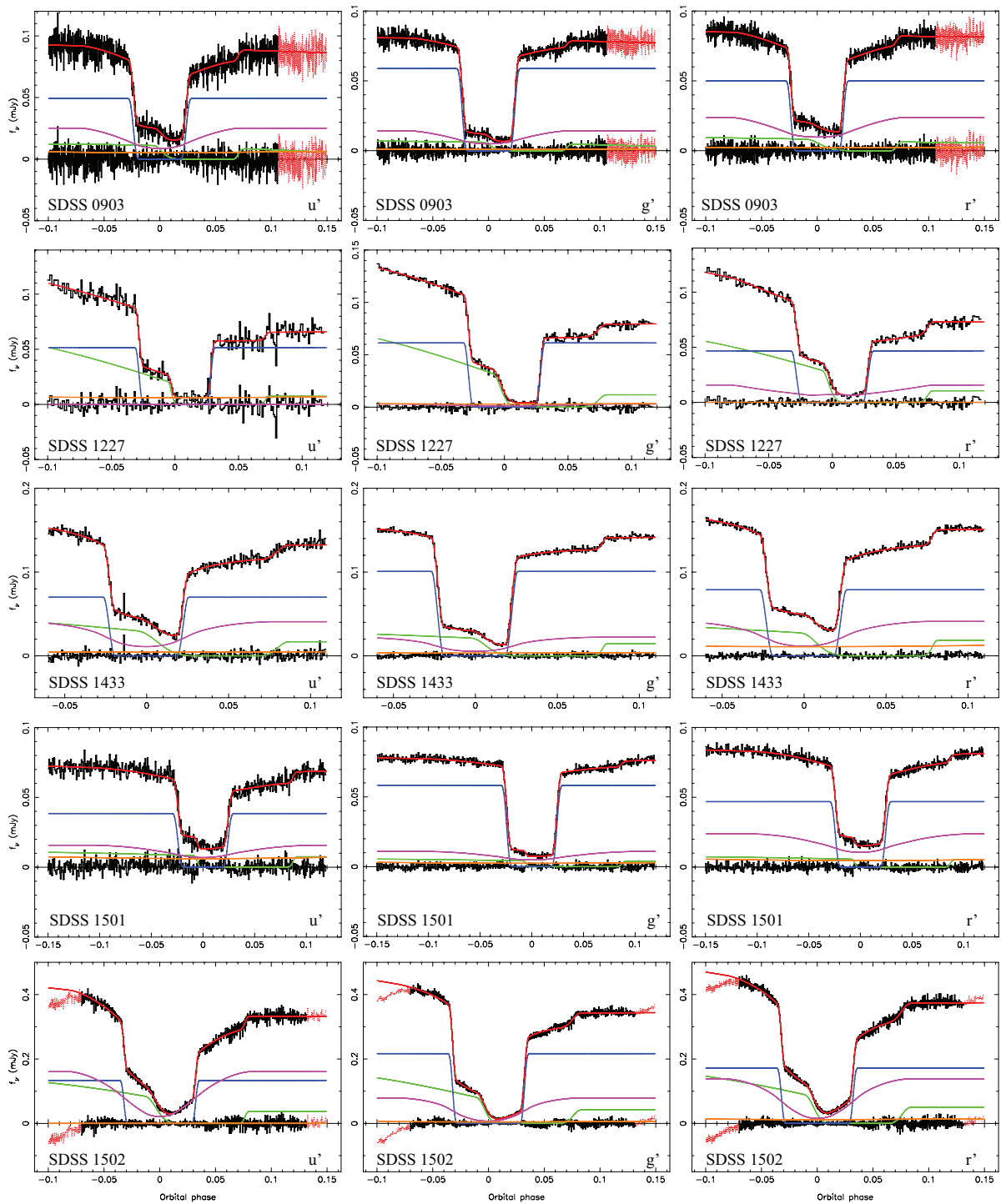} 
\caption{The phased-folded $u'g'r'$ lightcurves
  of target systems, fitted  using the model described in
  \protect\cite{littlefair07}. The data (black) are shown with the fit
  (red) overlaid and the residuals plotted below (black). Below are
  the separate lightcurves of the white dwarf (blue), bright spot
  (green), accretion disc (purple) and the secondary star
  (orange). Data points omitted in the fit are shown in red (light grey). }
\label{fig:model}
\end{center}
\end{figure*}

The model described in \cite{littlefair07} was fitted independently to the $u'g'r'$ lightcurves of each CV, and a weighted mean used to determine the final system parameters, which are shown in table~\ref{tab:params}. The only exception to this method was SDSS 1501. The ingress and egress features in the eclipse of SDSS 1501 are extremely faint. For this system we fit to the $g'$- and $r'$-lightcurves {\em simultaneously} to determine the system parameters. The resulting model was fit to the $u'$-lightcurve without optimisation of the system parameters in order to determine the depth of the white dwarf ingress/egress in the $u'$-band, and thus constrain the white dwarf temperature. The model fits to each lightcurve are shown in figure~\ref{fig:model}.

If we can increase our sample of donor star masses using observations from the literature it will enable more robust comparisons with, for example, theoretical evolutionary tracks. To this end, we searched the literature for short period objects which have mass determinations using the eclipse lightcurve technique. Two other systems with short orbital periods have reliable mass determinations derived from the eclipse lightcurves. These are XZ Eri \citep{feline04} and OY Car \citep{wood90}. The mass determination for XZ Eri uses a white dwarf mass-radius relation corrected to the appropriate temperature and is thus directly comparable with the results presented here. In contrast, the mass determination for OY Car used a zero-temperature analytical mass-radius relationship \citep{nauenberg72}. We have re-visited the mass determination in OY Car, combining the white dwarf radius and mass ratio estimates of \cite{wood90}, with the white-dwarf mass-radius relationship used here and in \cite{littlefair07} to obtain new system parameters. We assumed a limb-darkening parameter of 0.5, and set the uncertainty in the white dwarf radius to one-half the difference in radius estimates at limb-darkening parameters of 0.0 and 1.0. The white dwarf temperature at the time of the observations presented in \cite{wood90} is unknown.  We assumed a white dwarf temperature of 16500\,K, which is representative of OY Car in quiescence \citep{horne94}. Any error in our assumed white dwarf temperature will introduce a small corresponding error in the component masses for OY Car. The magnitude of this uncertainty is approximately 10 per cent for an error in temperature of 5000\,K. The new system parameters for OY Car are presented in table~\ref{tab:params}. The revised values represent a minor upwards revision from the values presented in \cite{wood90}; our estimates of the component masses are approximately equal to the upper limits in that paper.

\begin{table*}
\begin{tabular}{lcccccccc}
\hline
& SDSS 0903 & SDSS 1035$^1$ & SDSS 1227 & SDSS 1433 \\
\hline
Inclination $i$ & $80.8 \pm 0.1$ & $83.1 \pm 0.2$ & $83.9 \pm 0.2$ & $84.2 \pm 0
.2$\\
Mass ratio $q=M_{{r}}/M_{{w}}$ & $0.117\pm 0.003$ & $0.055 \pm 0.002 $ & $0.118 
\pm 0.003$ & $0.069 \pm 0.003$ \\
White dwarf mass $M_{{w}}/M_{\odot}$ & $0.96 \pm 0.03$ & $0.94 \pm 0.01$ & $0.81
 \pm 0.03$ & $0.868 \pm 0.007$ \\
Secondary mass $M_{{r}}/M_{\odot}$ & $0.112 \pm 0.004$ & $0.052 \pm 0.002$ & $0.
096 \pm 0.004$ & $0.060 \pm 0.003$\\
White dwarf radius $R_{{w}}/R_{\odot}$ & $0.0086 \pm 0.0003$ & $0.0087 \pm 0.000
1$ & $0.0103 \pm 0.0003$ & $0.00958 \pm 0.00008$ \\
Secondary radius $R_{{r}}/R_{\odot}$ & $0.141 \pm 0.003$ & $0.108 \pm 0.003$ & $
0.140 \pm 0.003$ & $0.109 \pm 0.003$ \\
Separation $a/R_{\odot}$ & $0.652 \pm 0.006$ & $0.622 \pm 0.003$ & $0.645 \pm 0.
007$ & $0.588 \pm 0.002$\\
White dwarf radial velocity $K_{{w}}/{\rm km\;s^{-1}}$ & $58 \pm 2$ & $29 \pm 1$ & $
54 \pm 2$ & $35 \pm 2$\\
Secondary radial velocity $K_{{r}}/{\rm km\;s^{-1}}$ & $494 \pm 5$ & $520 \pm 3$ & $
461 \pm 5$ & $511 \pm 2$\\
Outer disc radius $R_{{d}}/a$ & $0.27 \pm 0.02$ & $0.362 \pm 0.003$ & $0.292 \pm
 0.003$ & $0.358 \pm 0.001$\\
White dwarf temperature $T_{{w}}/\rm{K}$ & $13000 \pm 300$ & $10100 \pm 200$ & $
15900 \pm 500$ & $12800 \pm 200$\\ 
Distance (pc) & $274\pm10$ & $171\pm10$ & $380\pm10$ & $246\pm10$ \\

\hline
& SDSS 1501 & SDSS 1502 & SDSS 1507$^2$ & OY Car \\
\hline
Inclination $i$ &  $85.3 \pm 0.3$ & $88.9 \pm 0.8$ & $83.62 \pm 0.03$ & $83.3 \pm 0.2$ \\
Mass ratio $q=M_{{r}}/M_{{w}}$ & $0.067 \pm 0.003$ & $0.109\pm0.003$ & $0.0625 \pm 0.0004$ & $0.102 \pm 0.003$ \\
White dwarf mass $M_{{w}}/M_{\odot}$ & $0.80 \pm 0.03$ & $0.82\pm0.03$ & $0.91 \pm 0.07$ & $0.84\pm0.04$ \\
Secondary mass $M_{{r}}/M_{\odot}$ & $0.053 \pm 0.003$ & $0.090 \pm 0.004$ & $0.057 \pm 0.004$ & $0.086 \pm 0.005$ \\
White dwarf radius $R_{{w}}/R_{\odot}$  & $0.0104 \pm 0.0004$ & $0.0101\pm0.0004$ & $0.0091 \pm 0.0008$ & $0.0100 \pm 0.0004$ \\
Secondary radius $R_{{r}}/R_{\odot}$ & $0.108 \pm 0.004$ & $0.131\pm0.003$ & $0.097 \pm 0.003$ & $0.135 \pm 0.003$ \\
Separation $a/R_{\odot}$ & $0.589 \pm 0.008$ & $0.618\pm0.008$ & $0.54 \pm 0.01$ & $0.65\pm0.01$ \\
White dwarf radial velocity $K_{{w}}/{\rm km\;s^{-1}}$ & $33 \pm 2$ & $52\pm2$ & $34 \pm 1$ & $48\pm1$ \\
Secondary radial velocity $K_{{r}}/{\rm km\;s^{-1}}$ & $490 \pm 7$ & $479\pm6$ &  $550 \pm 14$ & $470\pm7$ \\
Outer disc radius $R_{{d}}/a$ & $0.452 \pm 0.009$  & $0.280\pm0.004$ &   $0.333 \pm 0.002$ & na \\
White dwarf temperature $T_{{w}}/\rm{K}$ & $12500 \pm 200$ & $12300\pm200$ &  $11000 \pm 500$ & 16500 (assumed) \\
Distance (pc) & $330\pm20$ & $170\pm20$ & $160\pm10$ & na \\
\hline
\multicolumn{4}{l}{$^1$ from \protect\cite{littlefair06}}\\
\multicolumn{4}{l}{$^2$ from \protect\cite{littlefair07}}
\end{tabular}
\caption{System parameters of the target CVs derived using
  lightcurve fitting.  $R_{{r}}$ is the volume radius of the
  secondary's Roche lobe \citep{eggleton83}.}
\label{tab:params}
\end{table*}

\begin{figure*}
\includegraphics[scale=0.5,angle=0]{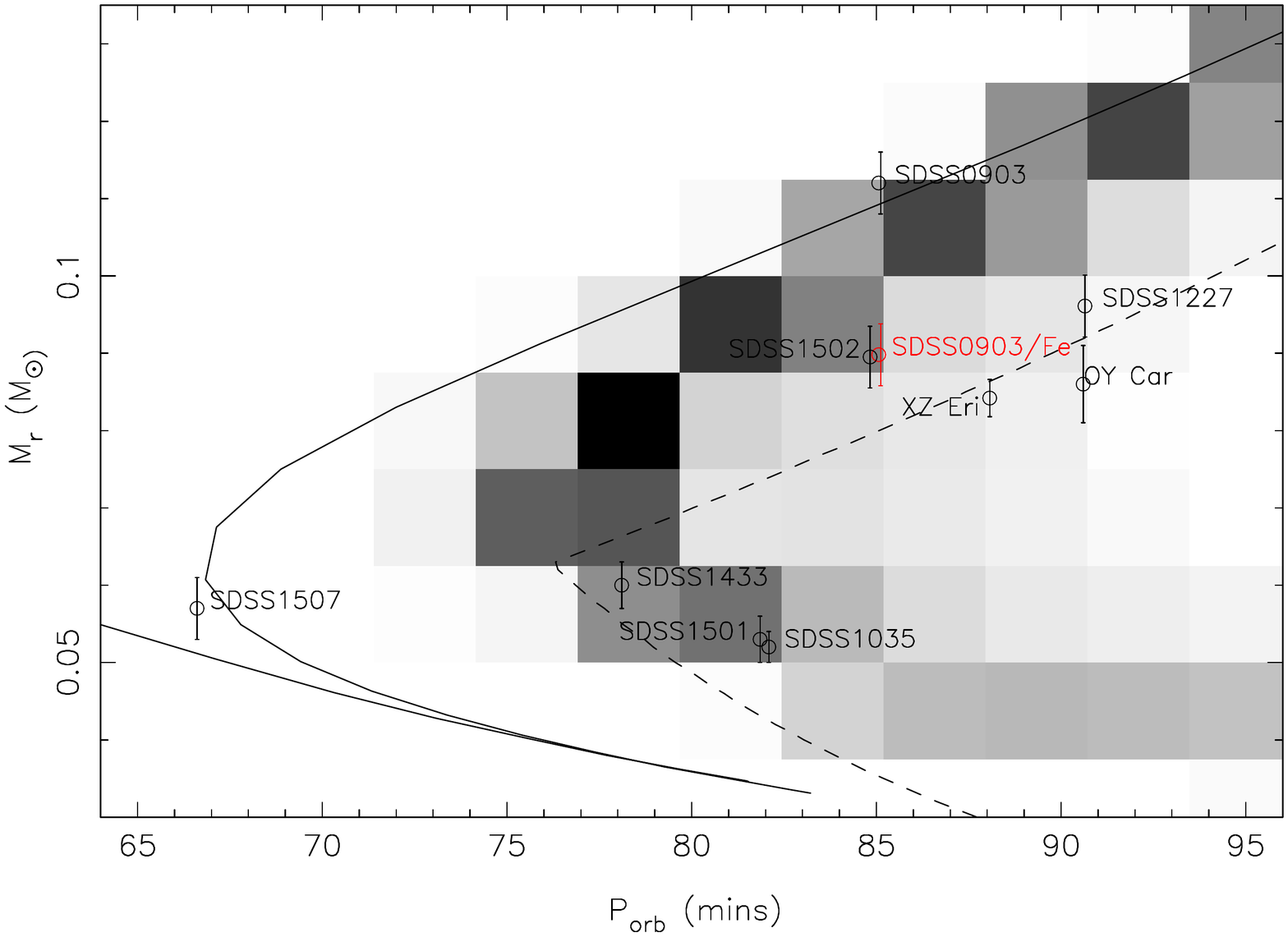} 
\caption{Donor star masses plotted against orbital period. The position of
SDSS0903 is shown twice; once assuming a CO-core white dwarf, and once with
an Fe-core white dwarf. Data for XZ Eri are taken from
\protect\cite{feline04}. The evolutionary tracks of
\protect\citet{kolb99} are shown as a solid line, whilst the
population density models of \protect\citet{willems05} (which include
enhanced angular momentum loss from circumbinary discs) are plotted in
grayscale. These models have been scaled to account for observational selecting
effects by weighting each point by L$_{\rm acc}^{1.5}$, where L$_{\rm acc}$ is the total accretion
luminosity of the system. The dashed line represents an empirical fit to donor star
masses derived using the superhump period-excess mass ratio
relationship \protect\citep{knigge06}. }
\label{fig:masses}
\end{figure*}

\begin{figure*}
\includegraphics[scale=0.5,angle=0]{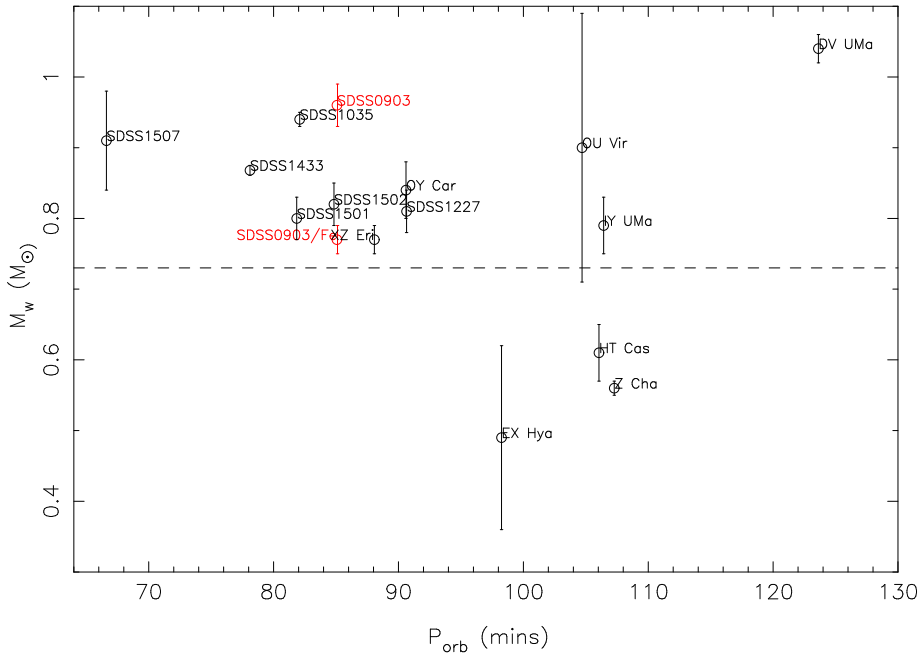} 
\caption{White dwarf masses as a function of orbital period. The position of
SDSS0903 is shown twice; once assuming a CO-core white dwarf, and once with
an Fe-core white dwarf. Data for those systems with system parameters which are not derived in this paper are taken from the compilation of reliable mass determinations in
\protect\citet{patterson05}. The mean mass for systems below the period gap (prior to our study -
\protect\citealt{knigge06}) is shown with a dashed line.}
\label{fig:wdmasses}
\end{figure*}

Including XZ Eri and OY Car, our analysis brings the total number of CVs with orbital periods below 95 minutes with mass determinations from the eclipse geometry to nine. Donor masses for these systems as a function of orbital period are shown in figure~\ref{fig:masses}. In figure~\ref{fig:wdmasses} we show the white dwarf masses for these nine systems, together with reliable mass determinations for longer period systems as compiled by \cite{patterson05}.

\subsection{Bright Spots}
\label{subsec:bs}
\begin{figure*}
\includegraphics[scale=0.5,angle=0]{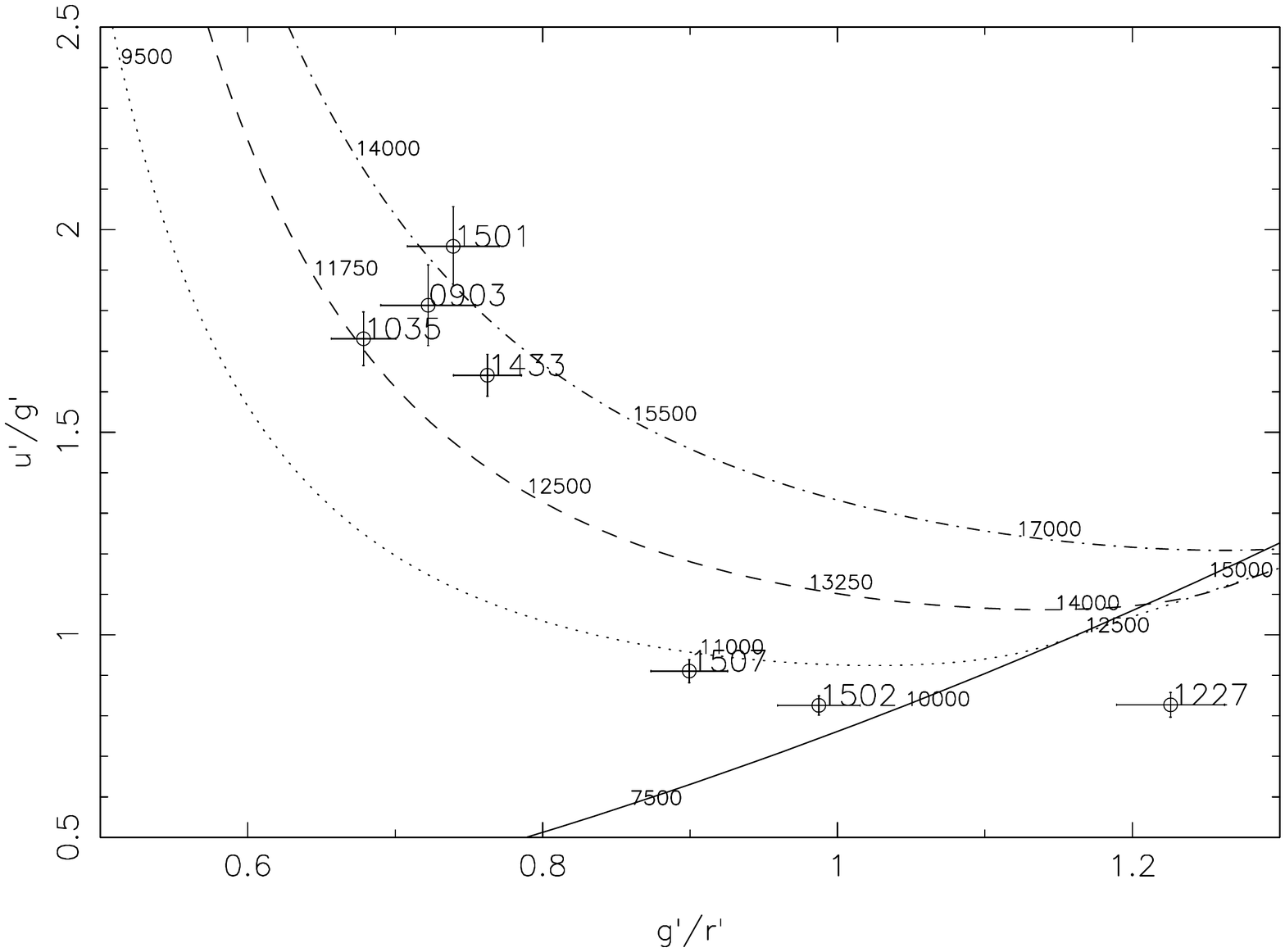}
\caption{Bright spot colours for our target systems. Also shown are the expected colours of black body emission (solid line),
and LTE Hydrogen slab models calculated using {\sc synphot} in {\sc iraf}.  Slab models with column densities of 
$10^{20}$ baryons/cm$^{-2}$ (dash-dotted line), $10^{21}$ baryons/cm$^{-2}$ (dashed line) and $10^{22}$ baryons/cm$^{-2}$ (dotted line) are shown. For each model, a temperature scale is plotted along the curve.}
\label{fig:bs}
\end{figure*}
As part of the model fitting we measure broadband $u'g'r'$ fluxes for the bright spots in these systems.  For most of the systems, the fluxes are largely determined from the magnitudes of bright spot ingress/egress. For those systems with a significant orbital hump (e.g SDSS 1227 and SDSS 1502), the size of the orbital hump also helps determine the bright spot fluxes.  Flux ratios for the bright spots are shown in figure~\ref{fig:bs}. Also shown are the expected flux ratios from black bodies of different temperatures, and from Hydrogen slabs in  local thermal equilibrium (LTE) at a range of temperatures and column densities. The Hydrogen slab models were calculated using {\sc synphot} in {\sc iraf}. Figure~\ref{fig:bs} shows that the bright spots in our systems fall into two groups: SDSS 0903, 1035, 1433 and 1501 have colours consistent with an optically thin Hydrogen slab, with temperatures between 12000 and 15000\,K; whilst SDSS 1227, 1502 and 1507 have colours more consistent with optically thick emission, and temperatures of 9000--11000\,K.

In section~\ref{subsec:mdot} we derive mass transfer rates from the white dwarf effective temperatures. Our white dwarf temperatures are constrained, in part, from the UV flux (see section~\ref{subsec:mdot} for full details), and so it is important to know if the bright spots contribute to the UV flux in these CVs. For the optically thin bright spots, it is unlikely that the bright spots contribute significantly to the emission bluewards of 2500\,\AA: the white dwarf is typically brighter in $u'$ than the accretion disc and bright spot combined, and the spectral energy distribution of an optically thin hydrogen slab drops rapidly towards the UV. In the case of the three systems with optically thick bright spots, however, extrapolating the black body emission into the UV suggests the bright spot contributes around one-third of the total flux at 2500\,\AA. Thus, for about half of our objects, the bright spot contributes a significant amount of UV flux.

\subsection{Mass-transfer rates}
\label{subsec:mdot}
\begin{figure*}
\includegraphics[scale=0.5,angle=0]{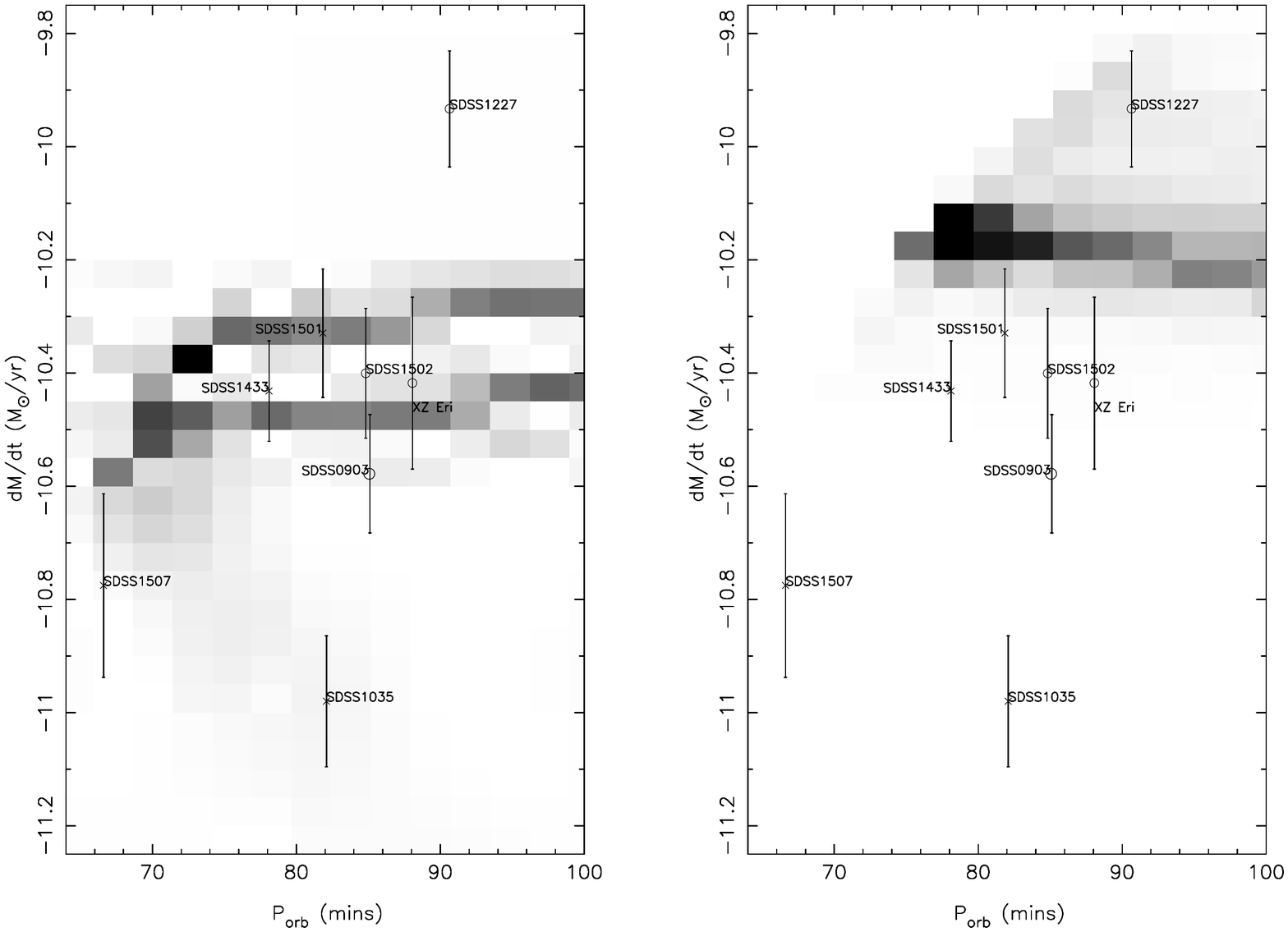}
\caption{Mass-transfer rates, as calculated from the white dwarf
effective temperatures (see text for details).  The population density models of
\protect\cite{willems05} are plotted in grayscale. The left-hand panel
shows models in which angular momentum loss is solely due to
gravitational radiation. The right-hand panel also includes additional
angular momentum loss from circumbinary discs. Pre-bounce systems
are denoted by open circles, post-bounce systems by asterisks.}
\label{fig:mdot}
\end{figure*}

Because we measure the broadband colours of white dwarf ingress/egress, we are able to obtain a crude estimate of the white dwarf effective temperature. The white dwarf temperatures in CVs are higher than expected given their age \citep{sion95}; a fact explained by compressional heating due to ongoing mass-transfer onto the white dwarf. The white dwarf temperature can thus be inverted to gain an estimate of the time-averaged accretion rate, albeit a model-dependent one \citep{townsley03}. \cite{townsley03} provide a relationship between the accretion rate and the {\em average} temperature over a nova cycle. We do not know where in the nova cycle our objects are, and so we include an additional uncertainty in the white dwarf effective temperature of 750K, as this is the typical variation in effective temperature through the nova cycle in short period systems (Townsley, priv.\,comm.). Accretion rates calculated in this manner are shown in figure~\ref{fig:mdot}, which shows that accretion rates are broadly in line with the expected values from gravitational radiation.  

Because our accretion rates depend critically on the white dwarf temperature, it is relevant to ask how robustly the white dwarf temperatures are determined by our method. One approach would be to calculate model fits to the SDSS spectra, and compare the white dwarf temperatures found. However, the optical data alone is not sufficient to constrain the white dwarf temperature. For example, in the case of SDSS 1035, models with white dwarf temperatures differing by more than 4000K gave acceptable fits to the optical data. Four of our objects have been observed by GALEX, however, and thus have available UV fluxes. These objects are SDSS 0903, 1035, 1501 and 1507. Following the prescription laid out in \cite{gaensicke06}, we fit models consisting of a red-star, white dwarf and LTE Hydrogen slab accretion disc simultaneously to the SDSS spectra and GALEX fluxes of these four objects.
For two of the objects (SDSS 0903 and 1501), plausible fits were found with the same white dwarf temperatures as determined from the lightcurve fitting. SDSS 1035 required a slightly hotter white dwarf (11400K compared with 10100K). For SDSS 1507, it was not possible to obtain a fit to the SDSS spectrum and GALEX fluxes simultaneously: the parameters determined from lightcurve fitting gave a plausible optical fit, but the UV fluxes are underpredicted by a factor of 3. We note that SDSS J1507 is one of the systems in which the bright spot probably contributes to the UV (see section~\ref{subsec:bs}). Our spectral fit does not include the bright spot, but it is unlikely that this could account for such a large discrepancy between optical and UV fluxes. A possible explanation is that SDSS 1507 was in a bright state at the time of the GALEX observations. Generally however, there is good agreement between white dwarf temperatures as derived from light curve fitting, versus those derived from spectral fits. We therefore believe that the effective temperatures presented here are accurate to $\sim 1000$\,K.

\section{Discussion}
\label{sec:disc}

\subsection{Post-period minimum cataclysmic variables}

A common feature of all population synthesis models for cataclysmic variables is the large numbers of systems which are predicted to have evolved past the orbital period minimum, and thus contain sub-stellar donor stars. The exact figure ranges from about 70 per cent \citep{kolb93,howell97} to  15--45 per cent \citep{willems05}. These predictions have always been in stark contrast with the {\em observed} population of CVs. \cite{littlefair03} reviewed the evidence for the existence of post-period minimum CVs, and found that there was no {\em direct} evidence that any system had evolved past the orbital period minimum. Since then, a small number of candidates have arisen in which there is reasonable {\em indirect} evidence for a sub-stellar donor \citep[e.g.][]{patterson05,howell06,burleigh06,araujo05}. However, none of these systems has a measured donor mass which is significantly below the hydrogen burning limit, and, as of November 2006, the observational dearth of post-period minimum CVs remained as significant as ever. 


The marked difference between the predicted and observed population of CVs is perhaps unsurprising. Post-period minimum CVs are difficult to identify as such, because the donor is so faint \citep{littlefair03}. In addition, it is possible that the discovery methods of CVs are strongly biassed against post-period minimum systems. Selection effects affecting the observed distribution of CVs are considered by \cite{knigge07} with particular attention to the Palomar-Green survey \citep{green86}. They find that existing samples of CVs are strongly biassed against short-period CVs, largely due to inadequate limiting magnitudes (see also \citealt{aungwerojwit06}). In addition, post-period minimum CVs will have long periods of quiescence, and may lack outbursts entirely \citep{kolb99}, making their discovery even more difficult. The SDSS can help overcome these problems. Whilst it too is a magnitude-limited survey  and thus still suffers considerable magnitude bias \citep{knigge07},  it is 2--3 mags deeper than previous surveys and should be sufficient to detect nearby faint post period-minimum CVs. CVs are identified in the SDSS by their spectral properties, having been flagged for spectroscopic follow up on the basis of their broadband colours (usually as candidate quasars). As a result, CVs in the SDSS need not be especially blue, need not have shown outbursts, and can be optically faint. The SDSS should therefore contain large numbers of post-period minimum CVs.  

To date, over 200 CVs have been discovered by the SDSS \citep{szkody02,szkody03,szkody04,szkody05,szkody06,szkody07}, of which nearly 100 have had orbital periods measured \cite[e.g.][]{gaensicke06,gaensicke08,southworth06,southworth07}. Remarkably, the period distribution of SDSS CVs reveals the long-sought ``period spike'' \citep{gaensicke08}, which strongly suggests that the SDSS is discovering post-bounce CVs in large numbers. A subset of the SDSS CVs show deep eclipses, making them excellent candidates for mass determination. Already, two SDSS CVs have been found to have unambigously sub-stellar donors \citep{littlefair06,littlefair07}. The work presented here brings this total to four (SDSS 1035, SDSS 1507, SDSS 1501 and SDSS 1433). Is this in line with theoretical expectations?  To give the most homogenous sample possible, we consider only the SDSS CVs with mass determinations.  SDSS 1507 is clearly unusual; with a period of 67-minutes, it is situated far below the well-defined orbital period minimum at 76.2 mins. SDSS 1507 most likely formed directly from a white dwarf/brown dwarf binary \citep{littlefair07}, or is a member of the old halo \citep{patterson08} and so we do not include it in our sample of post-period minimum CVs. This leaves three confirmed post-period minimum CVs amongst our small sample (SDSS 1035, SDSS 1501 and SDSS 1433). So far we have derived masses for seven SDSS CVs, leading to  an initial estimate that, amongst ``short-period'' SDSS CVs, $42\pm15$ per cent have evolved past the orbital period minimum, where ``short-period'' here means orbital periods below 95 minutes. \cite{willems05} find that between 15--40 per cent of CVs with orbital periods below 95 minutes should be post-bounce systems. The large range in their predictions encompasses differing assumptions about mass ratio distributions,  common envelope efficiency and the effect of circumbinary discs.  It is not possible to use our observations to constrain these assumptions without accounting for selection effects within the SDSS CV sample. A study of these selection effects is beyond the scope of this paper. One conclusion that can be reached, however, is that the fraction of post-bounce CVs within the SDSS sample is broadly consistent with that expected from population synthesis models; the missing post-bounce CVs have finally been found. Indeed, if we combine the results of this study with the observational confirmation of the long-predicted ``period spike'' \citep{gaensicke08}, we can see that the long-standing discrepancy between the observed and predicted CV population is beginning to be resolved in favour of the theoretical models.

Not all CVs with substellar donors are post-bounce systems. \cite{politano04} considers the formation of CVs directly from a detached white dwarf/brown dwarf binary, and finds that roughly 15 per cent of present day CVs should have formed in this manner. These systems should also be easy to detect amongst the CV population, as most of them will form with orbital periods below the observed minimum period of 76 minutes, and evolution to periods longer than 76 minutes is slow, taking 0.5--1.5\,Gyr. Furthermore, they should show similar mass-transfer rates, and thus similar observational properties to the post-bounce systems \citep{kolb99}, so they should also be detected by the SDSS. Of the $\sim$100 SDSS CVs with measured orbital periods, SDSS 1507 is the only one with a period below 76 minutes. Thus, the observed frequency of CVs forming directly from white dwarf/brown dwarf binaries is nearer 1 per cent than 15 per cent. The likely cause of this discrepancy is that brown dwarf companions to solar-type stars are roughly ten times less common than stellar companions \citep{grether06}. This is the well-known brown dwarf desert, and is not taken into account  in the calculations of \cite{politano04}. The scarcity of CVs forming directly from white dwarf/brown dwarf binaries is thus independent evidence for the brown dwarf desert.

\subsection{Donor star masses}
\label{subsec:masses}

If the population statistics of the SDSS CVs are broadly in line with expectations, figure~\ref{fig:masses} shows that the locus of the donor stars in the mass--orbital period plane are poorly reproduced by the theoretical models. At any given mass, the models of \cite{kolb99} significantly under-predict the observed period. As longer periods imply larger Roche lobes we can infer that the models of \cite{kolb99} underestimate the radii of the donor stars in short period CVs by roughly 10 per cent. The models of \cite{willems05}, in which the mass-transfer rate is enhanced due to the effect of circumbinary discs, do rather better in reproducing the observed donor masses, particularly for the post-bounce systems. Increasing the mass-transfer rate affects the locus of a system in the mass--orbital period plane because for low-mass donors the thermal timescale can be longer than the mass loss timescale. The effect of mass-transfer is thus to push the donor out of thermal equilibrium, leading to a donor that is larger than expected for a given mass. Because the donor star's thermal time-scale $t_{\rm KH} \sim GM^2/RL$ increases with decreasing mass, higher mass-transfer rates have a strong effect on the post-bounce CVs, but a minimal effect on the pre-bounce CVs. The effect of enhanced mass-transfer rates on the donor stars in CVs is well known, and enhanced mass-transfer rates are often invoked to explain the observed properties of short period CVs \cite[e.g.][]{patterson98,barker03}.

However, we encounter difficulties if we wish to explain the location of our systems in the P$_{\rm orb}$-M$_{\rm r}$ diagram via enhanced mass-transfer rates; the white dwarf temperatures are too cool to support this hypothesis. Figure~\ref{fig:mdot} shows the mass-transfer rates as inferred from the white   dwarf temperatures, compared to the expected mass-transfer rates from gravitational radiation alone, and the combined effect of gravitational radiation and circumbinary discs.  The mass-transfer rates, inferred from the white dwarf temperatures, are not consistent with enhanced mass-transfer rates, being roughly in line with the values expected from gravitational radiation alone. However, mass-transfer rates are notoriously difficult to estimate in CVs, so to what extent can we rely on the estimates presented here? It is certainly puzzling that, for example, the mass-transfer rates for SDSS 1501 and SDSS 1433 are so high, given their status as post-bounce systems.  These results might lead us to question both whether our determinations of mass-transfer rate are accurate, and whether the inferred mass-transfer rates truly reflect the long-term average.

Unlike mass-transfer rates from accretion luminosity, our mass-transfer rates, inferred from white dwarf temperatures, represent an average over $\sim$10$^4$\,yr \citep{townsley03}, and thus should be a better estimate of the long-term average mass-transfer rate.  However, a true estimate of the long-term average mass-transfer rates requires we average over $t  \sim t_{\rm KH} (H_r/R_r)$, where $H_r$ is the scale height of the donor's atmosphere. For the donors considered here, this timescale is $\sim$10$^5$\,yr: the mass-transfer rates presented here may still not reflect the long-term average rate. The estimate of mass-transfer rate depends both upon the white dwarf effective temperature and mass \citep{townsley03}. We are confident that the mass estimates are reliable (see section~\ref{sec:model} for a discussion), and the effective temperature measurements are probably reliable to $\sim 1000K$ (see section~\ref{subsec:mdot}). More accurate T$_{\rm eff}$ estimates could be obtained by estimating the white dwarf temperature from fits to UV spectra, but no such spectra exist for our objects.  Additionally, the mass-transfer rates presented here are, of course, model-dependent. The calculations of \cite{townsley03} assume that the white dwarf core reaches an equilibrium temperature, whereas detailed calculations following the white dwarf through many nova cycles suggest that the core may never reach equilibrium \citep{epelstain07}. Even if the white dwarf does reach equilibrium in the core, \cite{epelstain07} find significantly higher core temperatures than used in the calculations of \cite{townsley03}. More work on the question of mass-transfer rates in short period CVs is therefore highly desirable, both on theoretical and observational fronts.

\begin{figure*}
\includegraphics[scale=0.5,angle=0]{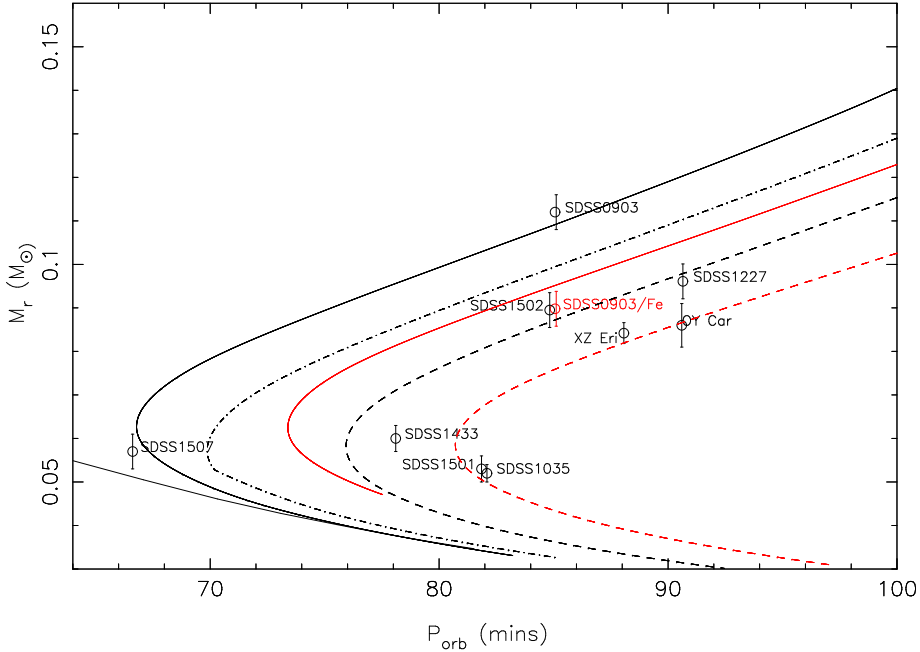}
\caption{Donor star mass versus orbital period, compared to theoretical predictions. The solid black line shows the ``standard'' sequence, with a 0.6 M$_{\odot}$ white dwarf primary \protect\citep{kolb99}. A second solid black line, entering the plot from the bottom left, shows a sequence which started mass-transfer as a white dwarf--brown dwarf binary. The other tracks show the effects of modifications to a ``standard'' sequence with a 1.0 M$_{\odot}$ white dwarf primary. The solid red (light grey) line shows the effect of including deformation of the donor, whilst the dashed red (light grey) line shows the effects of deformation and 50\% spot coverage combined, whilst the black dashed line shows the effect of 50\% spot coverage alone.  Lastly, the black dot-dashed line illustrates effect of inhibiting convection within the donor, showing a model in which the donor has a mixing length parameter of $\alpha=0.1$.}
\label{fig:theory}
\end{figure*}
If we take the mass-transfer rates shown in figure~\ref{fig:mdot} at face value, it implies that we cannot rely on enhanced mass-transfer to resolve the failure of models to reproduce the donor star mass--orbital period locus. Are there other possible explanations for the discrepancy? Processes linked to irradiation from the white dwarf are a possible explanation \citep{baraffe00}.
Our data, however, shows little evidence for this; there is no clear correlation between incident flux ($F_{\rm inc} \propto R_{\rm wd}^2 T^4_{\rm eff,wd} / a^2$) and the discrepancy in donor star radius between model and theory. Thus, whilst the statistics of our sample are too small to definitively rule out irradiation as a cause of the enhanced donor radii, we consider this unlikely. We expect that geometrical distortion due to tidal and rotational forces will have an effect on the radius of the donor. Hydrodynamical models which include distortion of the donor star \citep{renvoize02} show that this can provide an increase in radius of $\sim$5 per cent, although this may be reduced by a small amount due to thermal effects.  Thus distortion probably accounts for some, but not all, of the observed increase in donor radii. A final possibility is related to magnetic activity; the radii of low-mass stars, as determined from eclipsing binaries, are larger than predicted by some 10--15 per cent \citep[e.g][]{chabrier07,ribas06}.  \cite{chabrier07} hypothesize that this is explained by strong magnetic activity and/or rapid rotation which has a double impact, inhibiting convective efficiency and enhancing the presence of starspots. The effect is thus particularly prominent in eclipsing binaries because of the rapid rotation (P$<3$ days) of the binary components. The donor stars in CVs rotate more rapidly still, so it is not unreasonable to expect them to exhibit equally strong magnetic activity. However, because the donor stars in CVs are relatively cool and dense, we might expect the effects of starspots to dominate over the effects of inhibited convection. 

In figure~\ref{fig:theory}, we investigate the effects of distortion, inhibited convection and starspots on the predicted locus of systems in the mass--orbital period diagram. We begin with the models of \cite{kolb99}, and add modifications as necessary. Distortion is modeled following \cite{renvoize02}. Thermal effects are not included. The effects of inhibiting convection are investigated by altering the mixing length parameter of the donor star. In the absence of observational data on the temperatures of starspots in very low mass stars, we model the starspots as regions of zero emission, i.e. completely black spots \citep[see][for details]{chabrier07}. Figure~\ref{fig:theory} shows, as expected, that even a severe inhibiting of convection within the donor has a small impact upon the mass--orbital period relation. Similarly, it shows that distortion alone is insufficient to explain the location of our target systems within the donor mass--orbital period plane. If we include the effects of starspots, however, we can obtain a much closer agreement between observations and theory. The ``standard'' sequence with the addition of 50 per cent spot coverage does a good job of predicting the location of the pre-bounce systems, although it fails to describe the location of the post-bounce systems satisfactorily. However, all systems, excepting SDSS 0903 and SDSS 1507, are bracketed by the sequence including distortion and no starspots, and the sequence including distortion and 50 per cent spot coverage. SDSS 1507 is believed to have formed directly from a white dwarf--brown dwarf binary, and thus would not be expected to follow the sequences in figure~\ref{fig:theory}, whilst SDSS 0903 is discussed in section~\ref{subsec:wds}. Although the models shown in figure~\ref{fig:theory} have a simplistic treatment of the effects of distortion and starspots, we can see qualitatively that these two effects can in principle explain the location of our observed systems
in the P$_{\rm orb}$-M$_{\rm r}$ diagram. 

Is the presence of starspots on these cool donors a realistic assumption? Starspots are formed when energy transport by convection is inhibited by the local magnetic field. Thus, the stellar material must be sufficiently ionised for the magnetic field to influence the gas dynamics. A rough estimate suggests that this occurs for effective temperatures of 1600--1800\,K (see appendix~\ref{app:spots}). The models shown in figure~\ref{fig:theory} predict that all our target CVs have donors with effective temperatures in excess of 1700\,K. Therefore, it is not unreasonable to assume that starspots exist on the surface of these stars. Furthermore, although a spot coverage of 50 per cent may seem excessive, Doppler imaging of the donor stars in longer period CVs suggests spot coverage factors of 20--30 per cent \citep{watson07}, so it would seem that large spot filling factors are the norm for CV donors. We thus tentatively suggest that the radii of the donor stars in short period CVs can be explained by a combination of geometrical distortion and the effects of magnetic activity, particularly starspots. We note also that models including these effects correctly predict the observed minimum orbital period; it is likely that the two problems have the same solution.

Finally, we note that the empirical donor star mass-radius relationship, derived by \cite{knigge06} from the superhump period excess--mass ratio relationship, provides a good fit to the pre-bounce systems, but a poor fit to the donor stars in post-bounce systems (see figure~\ref{fig:masses}). This is not particularly surprising, as there are few systems near the period minimum suitable for calibrating the superhump period excess--mass ratio relationship. It would be highly desirable to monitor the post-bounce systems in the hope of detecting superhumps during outburst.

\subsection{White Dwarf Masses}
\label{subsec:wds}

At short orbital periods, low mass (M$\sim$0.5M$_{\odot}$) He-core white dwarfs (WDs) are expected to be common amongst CV primaries. Depending upon assumptions about the initial mass ratio distribution or common envelope efficiency, between 40 and 80 per cent of present-day CVs with orbital periods below 165 minutes are born with He-core WDs \citep{willems05}. High common envelope efficiencies favour He-core WDs; making it easier for them to survive the common envelope phase, whilst simultaneously increasing the likelihood that a CO-core WD system will emerge from the common envelope phase too widely separated to evolve into contact. Similarly, an initial mass ratio distribution which favours equal mass components will favour He-core WDs, as the majority of donor stars in CO-core WD systems will  be too massive for dynamically stable mass-transfer \citep{willems05}.

It is therefore extremely surprising that we find {\em no} low-mass, He-core WDs amongst our sample. In fact, figure~\ref{fig:wdmasses} shows that all of our white dwarfs are high in mass, being higher than the mean mass of 0.72\,M$_{\odot}$ for white dwarfs in CVs below the period gap \citep{knigge06}. The dominance of high mass white dwarfs within our sample is puzzling in the light of theoretical predictions and so we investigated if it could be due to selection effects. The SDSS CV survey looks for CVs in objects that have been selected for spectroscopic follow-up within the SDSS. The criteria used to select spectroscopic targets within SDSS are diverse, but objects which lie outside the stellar locus in colour space are likely to be selected as targets. For example, objects with $u'-g' \lta 0.45$ will be selected as quasar candidates. Thus $u'-g' \lta 0.45$ is sufficient (though not necessary) for an object to be flagged for spectroscopic follow-up.  Low-mass white dwarfs are redder than their high-mass counterparts and thus may not be selected for spectroscopic follow-up. To calculate the $u'-g'$ colours of CVs with low-mass white dwarfs, we assumed that the dominant contribution to the $u'$ and $g'$ light is the white dwarf. We further assumed a mass-transfer rate of $3\times10^{-11}$\,M$_{\odot}$\,yr$^{-1}$, as appropriate for a CV with an orbital period near 80 minutes, in which the sole source of angular momentum loss is gravitational radiation. Effective temperatures were calculated following \cite{townsley03} and the $u'-g'$ colour obtained from the models of \cite{bergeron95} for the appropriate white dwarf mass and effective temperature.  We find that CO-core white dwarfs
above $\sim$0.5\,M$_{\odot}$ should be sufficiently blue to be selected as spectroscopic targets within the SDSS. Given that the CV as a whole is likely to be bluer than the bare white dwarf, due to contributions from the accretion flow, it seems unlikely that selection effects in the SDSS survey can explain why all the white dwarfs in our sample are high mass (${\rm M} \ge 0.8 {\rm M}_{\odot}$). Nor it is likely that this tendency has been introduced when we selected systems from the SDSS CV sample for follow up (see introduction and discussion in \citealt{gaensicke08} for details) Thus, we must conclude that the dominance of high mass white dwarfs amongst the short period SDSS CV sample is a real effect.

A caveat to this statement must be issued; there is reason to suspect He-core white dwarfs may be cooler than a CO-core white dwarf of equivalent mass. Helium has a higher heat capacity/unit mass than Carbon, and Helium-core white dwarfs are larger at a given mass. We might then expect the Helium-core white dwarfs to be cooler than CO-core white dwarfs. Additionally, low-mass white dwarfs (with either a Helium or CO core) necessarily had lower mass companions at the onset of mass-transfer, or the ensuing mass-transfer would not be stable. Since Helium-core white dwarfs are generally of lower mass than CO-core white dwarfs it follows that, at a given orbital period, a CV with a Helium-core white dwarf has, on average, been accreting for less time than one with a CO-core white dwarf. This may mean that the Helium-core white dwarf has not had time to reach an equilibrium core temperature (\citealt{epelstain07} question whether even CO-core white dwarfs reach equilibrium core temperature in CVs). In summary, it is possible that a Helium-core white dwarf is cooler and redder than a CO-core white dwarf with equal mass and accreting matter at the same rate. Thus, whilst it is unlikely that selection effects can explain why the white dwarfs in our sample are so high in mass, it is not yet possible to say if selection effects are responsible for the absence of He-core white dwarfs amongst our sample. 

The dominance of high mass WDs within our sample has important consequences for the modelling of nova outbursts, and their effect on the long term evolution of CVs. Most calculations of the evolution of WDs under nova outbursts show a gradual {\em decrease} in the mass of the white dwarf. For example, the models of \cite{epelstain07} show a decrease in WD mass of approximately 5 per cent over 1000 nova cycles, whilst \cite{yaron05} find the erosion of the white dwarf mass is about 5 times smaller, for similar parameters.  The dominance of high mass white dwarfs in our sample of short period systems, which in turn are composed mostly of older CVs, means that any erosion of the WD in nova explosions must be minimal, or even that the WD might increase in mass with continuing nova outbursts.

Finally, we mention the peculiar system SDSS 0903. As seen from figure~\ref{fig:masses} if a CO-core white dwarf is assumed the donor star mass is much higher than in systems of similar orbital period. In contrast, if we adopt a mass-radius relationship appropriate for an Fe-core WD \citep{panei00}, the donor mass lies on the locus which is defined by the other systems. The existence of Fe-core white dwarfs is controversial; their only known formation route is a failed thermonuclear explosion of a degenerate white dwarf near the Chandrasekhar limit \citep{isern91}, and later calculations by the same authors showed that this formation route likely never occurs \citep{gutierrez96}.  Observational evidence for Fe-core white dwarfs first arose from Hipparcos measurements \citep{provencal98}, which showed that some white dwarfs (ProcyonB, EG50, and GD140) were unusually small for their measured masses, although the best candidate, ProcyonB, was later shown to be a normal white dwarf \citep{provencal02}. Recently, \cite{catalan07} have argued that an Fe-core composition for Hyades member WD0433+270 could make the cooling time for this object consistent with the Hyades cluster age, although this result is strongly dependent on the white dwarf cooling models used. We must therefore treat any claim of an Fe-core WD in SDSS0903 with extreme caution.  The alternative is that SDSS 0903 contains a CO-core WD, and the higher donor mass of 0.112M$_{\odot}$ is correct. This value lies very close to the ``standard'' evolutionary sequence of \cite{kolb99}, raising the possibility that the donors in CVs follow a range of tracks in the P$_{\rm orb}$-M$_{\rm r}$ diagram, with some systems following  the ``standard'' sequence and thus possessing ``normal'' donor stars. If this is the case, these systems must be quite rare, as objects following the ``standard'' sequence would populate the orbital period space below the observed minimum period. In fact,  the orbital period distribution of CVs in the SDSS shows a very sharp cutoff at the observed minimum period of 76 minutes \citep{gaensicke08}. Therefore, a donor star mass of 0.112M$_{\odot}$ implies an abnormal donor star. One possibility is that SDSS 0903 is metal-poor. Models of cool metal-poor stars show a deficit in opacity within the envelope, implying a smaller radius than a metal-rich star of equivalent mass \citep[e.g.][]{chabrier00}.
This explanation was invoked to explain the short orbital period of SDSS 1507 \citep{patterson08}. However, unlike SDSS 1507, SDSS 0903 has a small proper motion of 0.1\arcsec/yr \citep{rafferty01}, which is not easily reconciled with halo membership. Also, the observational evidence for metallicity dependent radii is not clear cut, with eclipsing binaries showing no correlation \citep{lopez-morales07}. Thus we are presented with two scenarios for this system, both of which seem unlikely a-priori; either the white dwarf or the donor is SDSS 0903 is unusual.  It is not possible to choose between these alternatives on the basis of current evidence; follow-up observations of SDSS 0903 are highly desirable. The system parameters quoted in table~\ref{tab:params} are those for a CO-core WD.

\section{Conclusions}
We present high-speed, three-colour photometry of a small sample of short period, eclipsing CVs taken from the Sloan Digital Sky Survey. We determine the system parameters via a parametrized model of the eclipse fitted to the observed lightcurve by $\chi^2$ minimization.  Three out of the seven systems possess brown dwarf donor stars and have thus evolved past the orbital period minimum. To the extent that our poor statistics and ignorance of selection effects will allow, this number confirms predictions that $\sim$40 per cent of CVs should have evolved past the orbital period minimum.  The donor star masses and radii are inconsistent with model predictions, with the majority of donor stars being $\sim$10 per cent larger than expected across the mass range studied here. One explanation for the discrepancy is enhanced angular momentum loss (e.g. from circumbinary discs), however the mass-transfer rates as  deduced from white dwarf effective temperatures are not consistent with enhanced mass-transfer rates. Alternatively, we
find the larger radii can be explained with a combination of the effects of geometrical distortion and starspots. Choosing between these explanations will require better estimates of the mass-transfer rates in these systems. This in turn will require refined estimates of the white dwarf temperature, together with additional work modeling the effect of accretion on the effective temperature of white dwarfs.

The white dwarfs in our sample show a strong tendency towards high masses. We show that this is unlikely to be due to selection effects, and instead is probably a real property of the short period SDSS CVs. The dominance of high mass white dwarfs within our sample implies that the white dwarfs in CVs are not significantly eroded by nova outbursts, or may in fact increase in mass over many nova cycles. Amongst our sample there are no He-core white dwarfs, despite predictions that 30--80 per cent of short period CVs should contain He-core white dwarfs. We are unable to rule out selection effects as the cause of this discrepancy. One white dwarf in our sample may be smaller than expected for a CO-core white dwarf, and possibly has an Fe-core, adding to the small number of Fe-core white dwarf candidates in the literature.

\section*{\sc Acknowledgments}
SPL acknowledges the support of a RCUK Fellowship. TRM acknowledges the support of a PPARC Senior Research Fellowship. CAW acknowledges the support of a PPARC Postdoctoral Fellowship.  ULTRACAM and SPL are supported by PPARC grants PP/D002370/1 and PPA/G/S/2003/00058, respectively. This research has made use of NASA's Astrophysics Data System Bibliographic Services. Based on observations made with the William Herschel Telescope operated on the island of La Palma by the Isaac Newton Group in the Spanish Observatorio del Roque de los Muchachos of the
Instituto de Astrofisica de Canarias. We thank Lars Bildsten for useful discussions.

\bibliographystyle{mn2e}
\bibliography{abbrev,refs,refs2}

\appendix
\section{Ionisation threshold for starspot formation}
\label{app:spots}
Starspots are formed when a flux tube penetrates the surface of the star. Because the ions in the flux tube are tightly bound to the field lines, and exert collisional forces on the material outside the flux tube, the flux tube acts as a barrier to the convection of material from outside it. This prevents warmer material from being convected to the point where the flux tube breaks the surface of the star. Hence the starspot is cooler, and darker, than the surrounding stellar surface.  However, for the starspot to form requires that the stellar atmosphere is sufficiently ionised for the magnetic field to influence the gas dynamics. If the gas is predominantly neutral, starspots cannot form. Our aim is to determine what level of fractional ionisation of the stellar atmosphere is required for starspots to form.
 
We can obtain a rule-of-thumb estimate by calculating the magnetic Reynolds number $R_m = v_{\rm conv} \delta r/\eta$, where $\eta$ is the magnetic diffusivity. From the atmosphere models of \cite{allard00}, $v_{\rm conv} \sim 10^4$ cm\,s$^{-1}$. We estimate the diameter of a flux tube to be 100\,km based upon the observed size of sunspots. If $R_m$ is much less than 1, the gas will be only loosely coupled to the field lines, and starspots will not form.  For $R_m  \ll 1$, we require $\eta \gg 10^{12}$\,cm$^2$s$^{-1}$. \cite{mohanty02} calculate $\eta$ for the atmosphere models of \cite{allard00}, and find that $\eta \gg 10^{12}$\,cm$^2$s$^{-1}$ at the surface of the star corresponds to an effective temperature of $T \gta 1800$\,K, with a fractional ionisation at the surface of $X \sim 10^{-10}$. This roughly tallies with observations of activity on low-mass stars, which finds activity starts to weaken around L0, or $T \sim 2500$\,K and is largely absent at L5, or $T \sim 1600$\,K \citep[e.g.][]{gizis00, reiners07}.

A slightly more analytical (but grossly simplified) approach is as follows. Consider a flux tube penetrating the stellar surface. Clearly, the point at which the surrounding gas can no longer penetrate the flux tube easily is the critical point for starspot formation. The dominant neutral species in a brown dwarf or very low mass star is H$_2$, whereas the dominant ion is Na$^+$, so that $n_{\rm e^-} = n_{\rm Na^+}$. The ions in the flux tube exert a drag force per unit volume

\begin{equation}
{\bf f_d} = \rho_{\rm H_2}\rho_{\rm Na^+} \left< \sigma v\right> ({\bf v}_{\rm Na^+} - {\bf v}_{\rm H_2})/ (m_{\rm Na^+}+m_{\rm H_2})
\label{1}
\end{equation}

where $\sigma$ is the cross-section for collisions between Na ions and H$_2$ molecules. We assume that the gas outside is dominated by neutral hydrogen atoms (i.e $\rho_{\rm total} \sim \rho_{\rm H_2}$), and we neglect the drag force from the electrons in the flux tube, as their mass is so much smaller than the mass of the ions. As long as the relative velocities of the two are smaller than the speed of sound then the Langevin approximation is valid, and $\left< \sigma v\right>_{\rm Na^+H_2} \sim 10^{-9}$\,cm$^3$\,s$^{-1}$ \citep{ciolek93}. We assume that, on average, the ions are stationary within the flux tube, and that the material outside moves with the convective velocity $v_{\rm conv}$. Hence the drift velocity between the two is $v_{\rm conv}$, which is 1--2 orders of magnitude below the sound speed under these conditions. Equation~\ref{1} then becomes 

\begin{equation}
{\bf f_d} \sim 10^{-9} \rho_{\rm H_2}\rho_{\rm Na^+} v_{\rm conv}/ (m_{\rm Na^+}+m_{\rm H_2})
\label{2}
\end{equation}

For the flux tube to be able to prevent the external gas from penetrating to the surface the work available from the drag force must exceed the kinetic energy of the external material, $f_d \delta r \gg \frac{1}{2} \left<\rho_{\rm ext} \right> v^2_{\rm conv}$. Because the external gas is dominated by neutral hydrogen $\left< \rho_{\rm ext} \right> \approx  \rho_{\rm H_2}$, hence
\begin{equation}
2 \times 10^{-9} \rho_{\rm H_2}\rho_{\rm Na^+} v_{\rm conv} \delta r/ (m_{\rm Na^+}+m_{\rm H_2}) \gg \rho_{\rm H_2} v^2_{\rm conv}
\end{equation}
$\rho_{\rm Na^+} = 23n_{\rm Na^+}m_H$, which gives
\begin{equation}
46 \times 10^{-9} n_{\rm Na^+}  m_H \delta r/ (23 m_{\rm H}+ 2 m_{\rm H}) \gg v_{\rm conv}
\end{equation}
or
\begin{equation}
2 \times 10^{-9}  n_{\rm Na^+} \delta r \gg v_{\rm conv}
\end{equation}
which gives
\begin{equation}
n_{\rm Na^+} \gg v_{\rm conv}/2\times10^{-9} \delta r.
\end{equation}
With $\delta r \sim 100$ km and $v_{\rm conv} \sim 10^4$ cm\,s$^{-1}$ this yields $n_{\rm Na^+} \gg 10^6$ cm$^3$.

Since the fractional ionisation, $X \approx n_{\rm Na^+} /n_{\rm H_2}$, we find that $n_{\rm H_2}X \gg 10^6$ cm$^3$. The models of  \cite{allard00} suggest that $n_{\rm H_2} \sim 10^{18}$ cm$^{-3}$, so we are left with $X \gg 10^{-12}$ as our result.

In conclusion, in order for us to have starspots forming, the fractional ionisation in the atmosphere needs to be greater than $10^{-10}$ to $10^{-12}$, which corresponds to effective temperatures of 1600-1800\,K \citep{mohanty02}.

\end{document}